\documentclass[aps,pre,twocolumn,showpacs]{revtex4}
\usepackage{bm}
\usepackage{amsmath}
\usepackage{amssymb}
\usepackage{graphicx}

\begin{document}

\title{Optimal estimate of probability density functions
from experimental data}

\author{R. Labb\'e}
\affiliation{Laboratorio de Turbulencia, Departamento de F\'isica,
Facultad de Ciencia, Universidad de Santiago de Chile, USACH.
Casilla 307, Correo 2, Santiago, Chile}

\date{June 1, 2006}

\begin{abstract}
A method providing optimal estimate of probability density functions
(PDFs) from time series is proposed. It allows almost arbitrary
resolution PDFs when applied to either, sampled analytic functions
or digitized data from experiments. When results are compared with
PDFs of the same data calculated using the standard histogram
method, a remarkable improvement is observed, especially in far
lateral regions of the PDF, where low probability events give poor
statistics.
\end{abstract}

\pacs{02.50.-r  05.10.-a}

\maketitle

Probability density functions (PDFs) are of main interest in
physical systems were the statistical description of magnitudes is
more appropriate than the detailed behavior in time and/or space of
one or more variables. In particular, in research in turbulence, it
is of interest to characterize properties with non-Gaussian
statistics, especially those related with small scale intermittency
\cite{Zhou,Stai}, responsible of slowly decaying wings in the PDF of
velocity differences at small scales, or the statistics of global
magnitudes like pressure \cite{Fau} or the injected power in
confined turbulent flows \cite{Lab1,Lab2} ---characterized by
non-symmetric PDFs showing an exponential or stretched exponential
wing on the left side. Being the events that contribute to these
particular features of the PDF rare, it is not often possible to
obtain a good statistics to accurately describe them, and their
effect on the the signals could appear to be rather marginal.
However, in view of their strength, they are detectable as a non
gaussian behavior in the tails of the PDF of the variable under
study, and given their importance in the description and
understanding of intermittency, among other effects, it is desirable
to have a reliable method to estimate the PDF of functions having
this kind of features. Additionally, when the PDF of short bursts in
a signal is being studied, it is worth to have a tool to estimate it
over relatively short time intervals. In these cases, the usual
method of building a histogram of the data is inadequate because the
number of points available could be not large enough. In this note I
propose a simple method to estimate the PDF of a sampled function,
like the data obtained when measuring the time evolution of some
quantity in an experiment, which produces remarkably good results.

The idea behind the method is simple: given a bounded time function
$f(t)$, $t\in [a,b] \subset \mathbb{R}$, with Fourier transform
$\mathcal{F}(\omega)$, a sampled version $f_{n} = f(t_{n})$ of
$f(t)$, with $t_{n}=nT$, $n=1,\ldots,N$, and $T$ the time interval
between samples, is accordingly with the sampling theorem, a
complete representation of the continuous function $f(t)$ provided
that: i) the function is band limited and ii) the highest frequency
contents in the spectrum of $f(t)$ is bounded by the Nyquist
frequency, defined as one half of the sampling frequency, i.e.

\begin{equation}
\mathcal{F}(\omega) = 0, \qquad  |w| > \pi/T. \label{sb}
\end{equation}

Thus, although the set of values $\{f_{n}\}$ is nothing but a
``small'' subset (one having zero-measure) of the set $\{f(t)|t\in
[a,b]\subset\mathbb{R}\}$, the Nyquist-Shannon-Kotelnikov sampling
theorem allows us to recover all the information contained in the
original function from the set of points $\{f_{k}\}$. When $f(t)$ is
defined for all $t \in (-\infty, \infty)$, the explicit expression
for its reconstructed version, $f_{r}(t)$, in terms of the samples
$\{f_{n}\}$ is

\begin{equation}
f_{r}(t) = \sum_{n=-\infty}^{\infty}f_{n} \frac{\sin
[\pi(t-nT)/T]}{\pi(t-nT)}. \label{fr}
\end{equation}

\noindent As we will see later, we do not need $f_{r}(t)$ in the
process of building the PDF. If we want to evaluate the PDF in $M$
points $y_{k}$, $k=1,\ldots,M$, a local approximation using few
samples near the points
$\{(t_{k},f(t_{k}))|f(t_{k})=y_{k},k=1,\ldots ,M\}$ will be enough.
Now, the usual method of binning the data to make a histogram, which
by appropriate normalization gives an estimate of the PDF of $f(t)$,
has the obvious drawback that only the values in the set $\{f_{n}\}$
are used. Thus, most of the information to build the PDF of $f(t)$
is lost. Alternatively, if we consider the continuous function
$f(t)$, it is intuitively obvious that the probability of finding a
certain value $\tilde y = f(\tilde t)$ in the interval $[y,y+\delta
y]$ should be proportional to the time $\delta t$ spent by the
function in traversing the arbitrarily small neighborhood
$[y,y+\delta y]$ of $\tilde y$. More precisely, $P(y)\propto |\delta
t/\delta y|\longrightarrow 1/|f'(t)|$ when $\delta y\longrightarrow
0$. As $f$ can take many times the value $y$ at many different
instants $t$, we need to add all these values together over the
whole time interval in which the PDF is being calculated. Then, we
have

\begin{equation}
P(y)=\frac{1}{N}\sum _{t:y=f(t)}\frac {1}{|f'(t)|}, \label{pdf}
\end{equation}

\noindent where N is a normalization constant, so that

\begin{equation}
\int _{-\infty} ^{\infty} P(y)dy=1. \label{norm}
\end{equation}

\noindent Equation \ref{pdf} can be seen as a particular case of
equation (5-5) in reference \cite{Pap}. Note that in (\ref{pdf}) the
set of values of $y$ can be constructed arbitrarily, provided that
$f(t)$ is defined for at least a subset of the values $y$ chosen to
evaluate $P(y)$. As a consequence, this allows ---as a by-product,
to increase arbitrarily the resolution in the evaluation of $P(y)$.

\begin{figure}[t]
\centering \vspace{-0.0cm} \hspace{-0.3 cm}
\includegraphics[width=.38\textwidth]{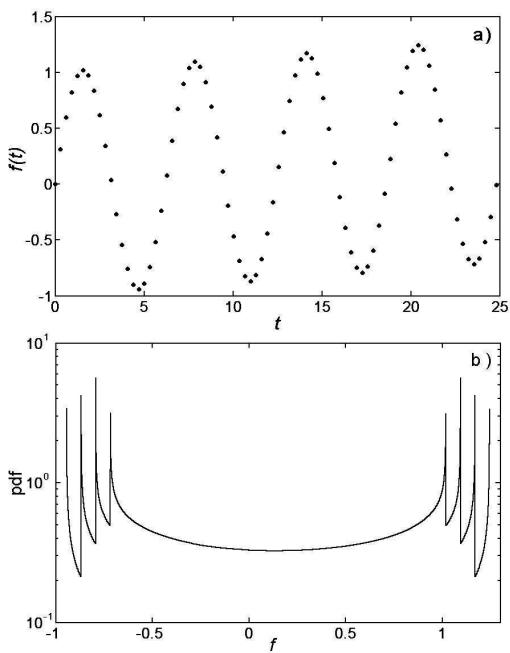}
\caption{{(a) Samples of a function from which the PDF is to be
calculated. (b) The resulting PDF, using $5000$ points.}}
\label{one}
\end{figure}

\begin{figure}[t]
\centering \vspace{-0.1cm} \hspace{-0.3 cm}
\includegraphics[width=.38\textwidth]{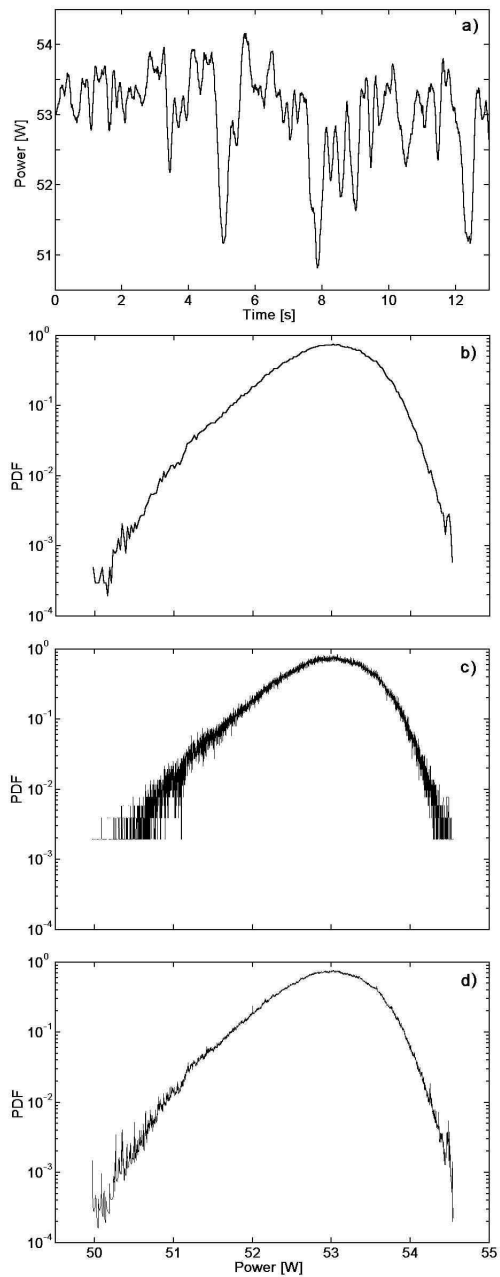}
\caption{{(a) Sample record of power injected in a confined
turbulent flow. (b) Its PDF estimate, obtained by binning the data
using $200$ bins. (c) A plot showing the result of increasing the
number of bins to $4000$. (d) Estimate of the PDF based on equation
3 (see text). The number of points is $4000$, as in the previous
plot.}} \label{two}
\end{figure}

It is important to mention here that a large number of points is not
strictly necessary for evaluating a PDF, although this certainly
helps in obtaining an accurate evaluation of the normalization
constant $N$. The reason is that formula (\ref{pdf}) corresponds to
the infinitely many points limit of the binning method (the proof is
straightforward). What is indeed needed is a good representation of
$f(t)$ in the neighborhoods of the zeroes of $f'(t)$, where the
r.h.s. of equation (\ref{pdf}) becomes singular. The advantage of
using a rather large number of points $y_i$ arises when the
normalization constant is calculated: given that $P(y)$ is singular
at the roots of $f'(t)$, then $N = \int \sum _{t:y=f(t)}
|f'(t)|^{-1}$ must be evaluated using a low order numerical
integration method. In the examples below, the trapezium rule was
used, which requires many points to give an accurate result. Another
concern is related to the singular values in the r.h.s. of equation
(\ref{pdf}). One would expect that zeroes in the denominator can
appear while running the numerical calculation. This is not the
case, because by picking an arbitrary value $y_k$, producing a set
$\{t_{kj}\}_{j=1,...,m_k}$ such that $f(t_{kj}) = y_k$, getting
$f'(t_{kj}) = 0$ for some $j$ is extremely unlike. To date, this has
never happened to me, neither in the examples given below nor in
other calculations. Thus, to compute a pdf, all we need is a
suitable, computationally efficient interpolation scheme to rebuild
the function $f$ near $(t,y)$, using some of the samples in the
discrete set $\{f_k \}$. Although this can be done with the help of
equation (\ref{fr}), using an interpolating polynomial through four
or six points around the point $(t,f(t))$ is far a better approach,
provided that the function corresponding to the samples is smooth
enough. When dealing with digitized data, this is ensured by the
anti-aliasing filter, except by the remaining electronic noise. I
will return to this point later.

To illustrate the method, let us start with the calculation of the
PDF of a few cycles of a sinus function plus a ``drift''

\begin{equation}
f(t) = \sin (2 \pi t/T) + \alpha t, \label{sndrf}
\end{equation}

\noindent with suitable values for the parameters, and using a
rather poor sampling. From the samples shown in figure 1 (a), and
using a third degree interpolating polynomial, the PDF displayed in
figure 1 (b) is obtained, using $5000$ point for $P(y)$. Note that
the expected singularities in this PDF are remarkably well
represented. Of course, there is no way to obtain this result by
binning the data shown in figure 1 (a).

As a second example, consider the figure 2 (a), displaying a $13$~s
sample of a $3000$~s length record of the power injected to maintain
the turbulence in a flow like those in references \cite{Lab1,Lab2}.
As this record was taken specifically to build the PDF of the
injected power, some oversampling was performed to allow numerical
smoothing on the data. In this case, a cutoff frequency of $50$~Hz
was used in the anti-aliasing filter, for a sampling rate of
$150$~sps (samples per second). The applied smoothing process is
such that the signal spectrum remains unchanged below the filter
cutoff frequency. With these two cautionary measures, it is possible
to use third order polynomials to locally reconstruct the signal
around the points $y_{k}$ chosen to build the PDF, using only four
neighboring samples. In figure 2 (b), a PDF built by using the
standard binning method is displayed. In this case, $200$ bins were
used. The PDF looks very acceptable, thanks to the length of $4.5
\times 10^{5}$ samples of the whole data record. If we want to
increase the accuracy of the PDF by increasing the number of bins,
things begin to go from bad to worse. Figure 2 (c) shows the PDF
obtained when $4000$ bins are used. Obviously, trying this is not
quite reasonable. However, by using the method that I propose here,
a remarkably good $4000$ point estimate of the PDF is effectively
obtained, as displayed in figure 2 (d). When compared with figure 2
(a), it is clear that all of the features present there are
recovered, but with a highly increased level of detail.

In conclusion, in this note I report a powerful method to estimate
the PDF of magnitudes obtained as time series from essentially
continuous functions of time, like those resulting by digitizing the
signal resulting from the output of an antialiasing filter ---a very
common experimental scenario. In contrast to the usual binning
procedure, the method I propose here can yield, in principle,
optimal accuracy and arbitrary resolution in the resulting estimate
of the probability density function, even for rather small data
sets.
\\
\\
This work benefited of the financial support provided by FONDECYT,
under project No. 1040291.


\begin{thebibliography}{99}

\bibitem{Zhou} T. Zhou, Z. Hao, L. P. Chua, and S. C. M. Yu, Phys. Rev. E {\bf 71}, 066307 (2005)
\bibitem{Stai} A. Staicu and W. van de Water, Phys. Rev. Lett. {\bf
90}, 094501 (2003)
\bibitem{Fau}  P. Abry, S. Fauve, P. Flandrin, and C. Laroche,  J. Phys. II {\bf 4} 725 (1994)
\bibitem{Lab1} R. Labb\'e, J.-F. Pinton and S. Fauve, J. Phys. II {\bf 6}, 1099
(1996)
\bibitem{Lab2} J.-F. Pinton, P. C. W. Holdsworth, and R. Labb\'e,
Phys. Rev. E {\bf 60}, R2452 (1999)
\bibitem{Pap} A. Papoulis, {\it Probability, random variables and stocastic processes},
(McGraw-Hill, New York, 1991), chap. 5

\end{thebibliography}
\end{document}